\documentclass[prb,showpacs,superscriptaddress,twocolumn]{revtex4-1}

\usepackage{graphicx}

\usepackage{graphicx}
\usepackage{amssymb}
\usepackage{bm}
\usepackage{amsmath,amsfonts,latexsym,color}

\def\beq{\begin{equation}}
\def\eeq{\end{equation}}
\def\beqn{\begin{eqnarray}}
\def\eeqn{\end{eqnarray}}

\begin{document}
\title{Scaling of electrical and thermal conductivities in an almost integrable chain}
\author{Y.\ Huang}
\affiliation{Department of Physics, University of California,
Berkeley, California 94720, USA}
\author{C.\ Karrasch}
\affiliation{Department of Physics, University of California,
Berkeley, California 94720, USA} \affiliation{Materials Sciences Division,
Lawrence Berkeley National Laboratory, Berkeley, California 94720, USA}
\author{J.\ E.\ Moore}
\affiliation{Department of Physics, University of California,
Berkeley, California 94720, USA} \affiliation{Materials Sciences Division,
Lawrence Berkeley National Laboratory, Berkeley, California 94720, USA}
\date{\today}


%

\begin{abstract}
Many low-dimensional materials are well described by integrable one-dimensional models such as the Hubbard model of electrons or the Heisenberg model of spins.  However, the small perturbations to these models required to describe real materials are expected to have singular effects on transport quantities: integrable models often support dissipationless transport, while weak non-integrable terms lead to finite conductivities.  We use matrix-product-state methods to obtain quantitative values of spin/electrical and thermal conductivities in an almost integrable gapless ($XXZ$-like) spin chain. At low temperatures, we observe power laws whose exponents are solely determined by the Luttinger liquid parameter. This indicates that our results are independent of the actual model under consideration.
\end{abstract}

\maketitle

The physics of many one-dimensional systems with idealized interactions is rather special: the quantum Hamiltonian has infinitely many independent conserved quantities that are sums of local operators.  Such Hamiltonians are called ``integrable'' in analogy with classical Hamiltonian systems that decompose into independent action and angle variables.  Examples relevant to experiments on crystalline materials include the Hubbard model of electrons and the XXZ model of spins; ultracold atomic systems can realize integrable continuum models of bosons.  However, in all these cases it is expected that integrability is only an approximation to reality and that experimental systems have integrability-breaking perturbations which, while small, drastically change some of the physical properties.

Transport properties provide an experimentally important example of the effects of non-integrable perturbations. In integrable systems, parts of charge, spin, or energy currents are conserved, and thus transport is dissipationless even at non-zero temperature. This corresponds to a finite ``Drude weight'' $D$ in the frequency-dependent conductivity:~\cite{betheT0,bethezotos,bethekluemper,edfabian,edmillis,qmcsorella,qmcgros,Sirkerprl,Sirker,prosen,karrasch,qmctherm,bethetherm,fabianrev,cher1,sirkerw}
\beq
\sigma(\omega) = 2 \pi D \delta(\omega) + \sigma_\textnormal{reg}(\omega)~.
\eeq
In reality, many quasi-one-dimensional systems are expected to be well described by integrable Hamiltonians \textit{plus weak non-integrable perturbations}. The zero-frequency conductivity is regularized ($D=0$) by these perturbations \cite{integrab,edmillis,qmcsorella,qmcgros,roschandrei,Jungprb,Jung,Jungth,fabianrev,edfabian,qmctherm}. An experimental example is the large (but not dissipationless) anisotropic thermal transport observed in  Sr$_{14}$Cu$_{24}$O$_{41}$ attributed to a long mean free path of quasi-1D magnons \cite{hess,ott,hlubek}. However, computing $\sigma_\textnormal{reg}(\omega)$ quantitatively for a microscopic non-integrable Hamiltonian is a challenging problem.

We study a generic gapless non-integrable system (a $XXZ$-like spin chain) using density matrix renormalization group (DMRG) methods, which were developed in the past few years to access finite-temperature dynamics of correlated systems. Using linear prediction, we extrapolate current correlations functions to large times. This allows to quantitatively observe the destruction of the thermal and electrical Drude weights; we compute the corresponding conductivities and analyze how they depend on temperature and the strength of the non-integrable perturbation. Our key observation is power law scaling behavior at low temperatures. The corresponding exponents are functions of the equilibrium Luttinger liquid parameter, which indicates that our results should hold for any gapless non-integrable model in which no conserved quantity protects the current. We compare our numerics to a low-energy field theory calculation using bosonization techniques which we obtain by adapting previous results \cite{Sirker,Sirkerprl} to the non-integrable perturbation in our model.

\begin{figure}[b]
\includegraphics[width=0.7\linewidth]{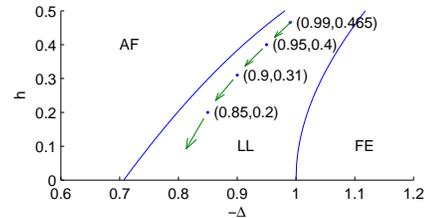}
\caption{The phase diagram \cite{Alcaraz,Okamoto} of (\ref{XXZ}). The points share the same Luttinger liquid parameter $K\approx2.4$ computed by DMRG \cite{llpaper}.}
\label{phase}
\end{figure}


\begin{figure}
\includegraphics[width=0.7\linewidth]{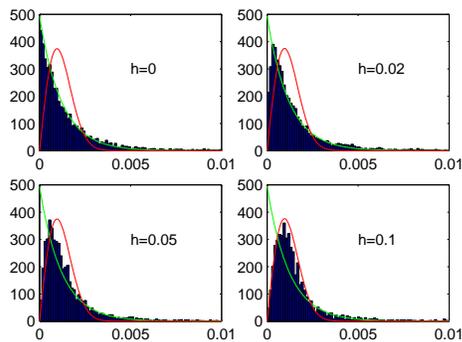}
\caption{The distributions of $\{E_{i+1}-E_i\}$, where $E_1\le E_2\le\ldots\le E_{4862}$ are the eigenenergies of (\ref{XXZ}) for $L=18, \Delta=-0.8$ in the $S_z=1,k=2\pi/9$ sector (the total magnetization $S^z=\sum_{i=1}^LS_i^z$ and the momentum $k$ are conserved).  The green curve and red curve are best-fit exponential and Wigner-Dyson (orthogonal ensemble) distributions respectively.  Neither the Poisson nor the Wigner-Dyson distribution appears clearly if we do not restrict to a symmetry sector of the model.  The crossover to Wigner-Dyson with increasing $h$ is observed independent of the choice of $\Delta$ and symmetry sector.}
\label{stat}
\end{figure}

{\it The model}.  We use the XXZ model in presence of a staggered magnetic field which breaks integrability (which we will demonstrate explicitly by considering level statistics). Its Hamiltonian is given by $H=\sum_{i=1}^L h_i$ with
\begin{equation}
h_i = S_i^xS_{i+1}^x+S_i^yS_{i+1}^y+\Delta S_i^zS_{i+1}^z+\left(-1\right)^ihS_i^z .
\label{XXZ}
\end{equation}
Without the staggered field, (\ref{XXZ}) is gapless for $|\Delta|\le1$ and gapped for $|\Delta|>1$. Bosonization leads to the low-energy effective Hamiltonian for $|\Delta|\le1$ and infinitesimal $h$:
\begin{equation}\label{effective}\begin{split}
H=& \frac{v}{2}\int dx\left(\Pi^2+\left(\partial_x\phi\right)^2\right)+ch\int dx\cos\left(2\sqrt{\pi K}\phi\right)\\[1ex]
& + H_\textnormal{umklapp} + H_\textnormal{band curv.} + H_{\textnormal{higher terms in } h}~,
\end{split}\end{equation}
where $\Pi$ is the conjugate momentum of the bosonic field $\phi$ with the canonical commutation relation $[\phi(x),\Pi(y)]=i\delta(x-y)$. The first term in (\ref{effective}) describes a Luttinger liquid. The Luttinger liquid parameter $K$ is given through Bethe ansatz: $2K\arccos(-\Delta)=\pi$, and the coefficients $v$ and $c$ are also known exactly \cite{Lukyanovprb,Lukyanov}. As the scaling dimension of $h$ is $2-K$, the second term in (\ref{effective}) is relevant and opens a gap for $K<2$ or $-\sqrt2/2<\Delta\le1$; the term is irrelevant and (\ref{XXZ}) remains in the gapless Luttinger liquid phase for $K>2$ or $-1<\Delta<-\sqrt2/2$ (Fig.~\ref{phase}). In the regime where $h$ is relevant, its effects have been studied perturbatively \cite{affleck,affleck2} and compared to experiments on spin diffusion in  copper benzoate \cite{hsexp,hsexp2}.

Integrability is well-defined in classical mechanics, but the definition
of its quantum counterpart remains a subject of debate \cite{Caux}. It is
generally believed that the level spacing distribution (the distribution of
the differences of the adjacent eigenenergies) is the exponential
distribution for an integrable model, as levels appear as a Poisson
process, and the Wigner-Dyson distribution for a nonintegrable model. 
Intuitively, two nearby levels in an integrable model likely have different
values of at least one integrable quantity, and thus live in different
sectors of Hilbert space that are independent of each other; hence their
energies are uncorrelated.  Non-integrable models do not have an extensive
number of such sectors and show energy level repulsion.  The belief has
been verified numerically on a variety of models \cite{Rabson, Santos}. We
perform an exact diagonalization of (\ref{XXZ}) with periodic boundary
conditions. Fig. \ref{stat} shows the level spacing distributions, and the
crossover from Poissonian behavior at $h=0$ to the Wigner-Dyson
distribution at nonzero $h$ is clear. Hence (\ref{XXZ}) is nonintegrable
for nonzero $h$.


{\it Numerical approach}. The DC charge (c) and heat (h) conductivities can be computed via the Kubo formula
\begin{equation}
\sigma= \lim_{t_M\to\infty}\lim_{L\to\infty} \frac{1}{LT}\,\textnormal{Re} \int_0^{t_M} \langle J(t)J(0)\rangle\, dt~,
\label{Kubo}
\end{equation}
where the corresponding current operators $J=\sum_{i=1}^L j_i$ are defined through a continuity equation \cite{edfabian}:
\begin{equation}\begin{split}
\partial_t h_i = j_{\textnormal{h},i}-j_{\textnormal{h},i+1}& ~\Rightarrow~  J_\textnormal{h} = i \sum_{i=2}^L [h_{i-1},h_i]~,\\
\partial_t S_i^z = j_{\textnormal{c},i}-j_{\textnormal{c},i+1}& ~\Rightarrow~  J_\textnormal{c} = i \sum_{i=2}^L [h_{i-1},S_i^z]~.
\end{split}\label{current}\end{equation}
The current correlation functions $\langle J(t)J(0)\rangle$ can be computed efficiently using the real-time finite-temperature density matrix renormalization group (DMRG) algorithm \cite{dmrgrev,white,frank,white2,vidal,daley,frank2,mettts,barthel1,barthel2} introduced in \cite{karrasch}. DMRG is essentially controlled by the so-called discarded weight $\epsilon$ (we ensure that $\epsilon$ is chosen small enough and that $L$ is chosen large enough to obtain numerically-exact results in the thermodynamic limit). The simulation is stopped when the DMRG block Hilbert space dimension $\chi$ reaches $\chi\sim1000-1500$. This allows to access time scales $t\sim 10-20$ which are larger than the inherent microscopic scale $t=1$.

\begin{figure}[t]
\includegraphics[width=0.95\linewidth,clip]{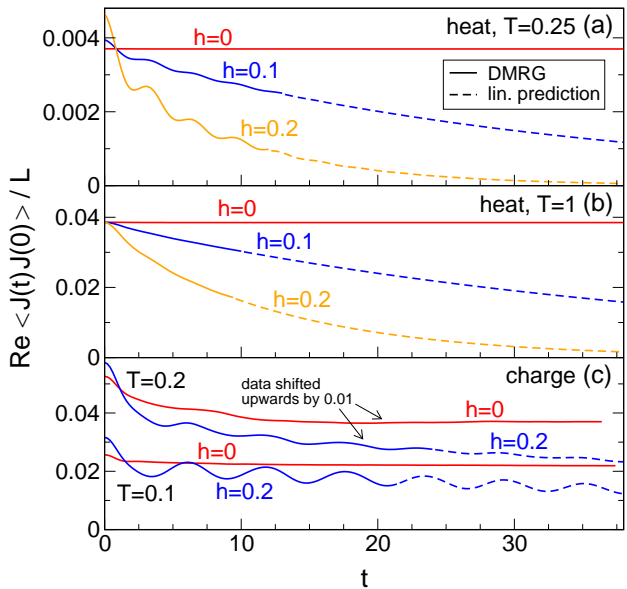}
\caption{Heat and charge current correlation functions $\langle J(t)J(0)\rangle$ at $\Delta=-0.85$. Their Fourier transform determines the corresponding conductivities through Eq.~(\ref{Kubo}). Our data is consistent with the following picture: The Drude weight (Eq.~(\ref{eq:ecorr})) is nonzero only in the integrable case $h=0$; a nonintegrable perturbation $h>0$ renders the conductivity finite. DMRG data (solid lines) is extrapolated using linear prediction (dashed lines). }
\label{corr}
\end{figure}

Results for $\langle J(t)J(0)\rangle$ are shown in Fig.~\ref{corr}. In the integrable case $h=0$, the heat and charge Drude weights
\begin{equation}\label{eq:ecorr}
D = \lim_{t\to\infty}\lim_{L\to\infty} \frac {\textnormal{Re } \langle J(t)J(0)\rangle}{2LT}
\end{equation}
are finite. A nonzero $h>0$ renders the model nonintegrable; one expect that the current correlation functions decay to zero at large times and that the conductivities become finite. Our data is consistent with this. In order to compute the integral in Eq.~(\ref{Kubo}) quantitatively, $\langle J(t)J(0)\rangle$ needs to be extrapolated. The heat current correlation function at intermediate to large temperatures $0.5\leq T\leq\infty$ (see Fig.~\ref{corr}(b)) can be fitted by a single exponential function $\exp(-\lambda t)$ \footnote{An estimate of the Drude weight has been given for this model~\cite{lima} under the assumption that thermal currents do not decay when $h > 0$; as shown in Fig. 3, we find that thermal currents decay and the Drude weight is zero.}. Oscillations develop at small $T$, but it is reasonable to assume that $\langle J(t)J(0)\rangle$ can be described by sums of exponentially decaying terms $\exp(-\lambda_n t +i\omega_nt)$ (the same holds for the charge current correlation function). This motivates us to use so-called linear prediction \cite{dmrgrev,barthel1,linpred} as an extrapolation procedure. Its stability can be tested by varying fit parameters (e.g.~the number of terms or the fit interval) and by checking sum rules (see below); we can obtain accurate results for the heat conductivity at any $h$ and temperatures $0.2\lesssim T\leq \infty$ as well as for the charge conductivity at intermediate to large $h$ and small $T\lesssim 0.3$.

\begin{figure*}[t]
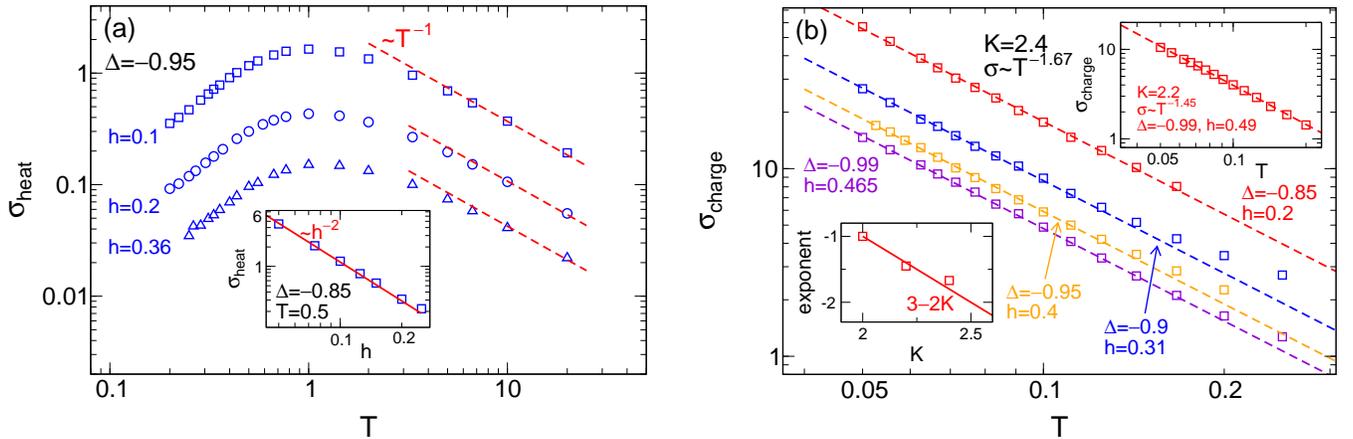

\includegraphics[width=0.475\linewidth,clip]{heat.eps} \hspace*{0.025\linewidth}
\includegraphics[width=0.475\linewidth,clip]{charge.eps}
\caption{Scaling of the heat and charge conductivities of a nonintegrable gapless spin chain. At low temperatures, $\sigma_c(T)\sim T^{\alpha_c(K)}$ is governed by power laws whose exponents are a universal function of the Luttinger liquid parameter $K$ only. At fixed $T$, $\sigma$ diverges as $h^{-2}$ for $h\to0$ with $h$ being the strength of the nonintegrable perturbation \cite{Jungth}. Note that our definition of $\sigma_h$ via Eq.~(\ref{Kubo}) differs from the most common one \cite{fabianrev} by a factor $T^{-1}$. The conductivity data points shown have a statistical error of at most 5\%. This error results from the extrapolation of $\langle J(t)J(0)\rangle$ via linear prediction and can be estimated by carrying out linear prediction using a variety of fit parameters. The error associated with the error of the raw DMRG data for $\langle J(t)J(0)\rangle$ is negligible compared to this.}
\label{sigma}
\end{figure*}

{\it Scaling of the conductivities: numerical results}. DMRG data for the heat and charge conductivities is shown in Fig.~\ref{sigma}. For fixed $T$ and small $h$, one expects $\sigma$ to diverge as \cite{Jungth}
\begin{equation}
\sigma \sim h^{-2}~~\textnormal{for}~~h\to0~. 
\end{equation}
This is consistent with our results (for the thermal case see the inset to Fig.~\ref{sigma}(a); data for $\Delta=-0.95$ is similar). At small $T$, $\sigma_c$ features power laws with nontrivial exponents:
\begin{equation}
\sigma_c \sim T^{\alpha_c} ~~\textnormal{for}~~T\to0~. 
\end{equation}
Our model is a Luttinger liquid at low energies and one thus expects $\alpha_c$ to be a universal function of the Luttinger liquid parameter $K$ only. This is verified in Fig.~\ref{sigma}(b) which shows $\sigma_c(T)$ for different parameter sets $(\Delta,h)$ which share the same $K$ ($K$ is a continuous function of $h$ and $\Delta$; it can be obtained from an independent ground state DMRG calculation,\cite{llpaper} and the parameters in Fig.~\ref{fig:sigma}(b) are determined numerically such that they all correspond to the same $K$). The exponent $\alpha_c(K)$ varies strongly with $K$; it is consistent with the analytic predition $\alpha_c=3-2K$ established below (see the insets to Fig.~\ref{sigma}(b)). The heat conductivity is only accessible for intermediate temperatures $T\gtrsim 0.2$; our data in this regime is almost independent of $K$ (see Fig.~\ref{sigma}(a)), suggesting that we have not reached the limit of low $T$.

\begin{figure}[b]
\includegraphics[width=\linewidth,clip]{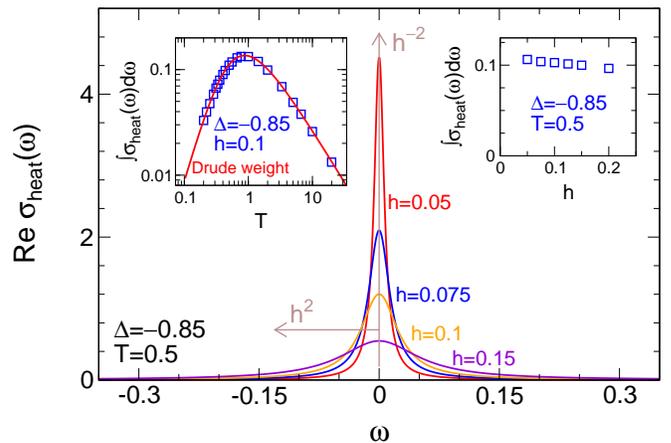}
\caption{AC heat conductivity. A Drude peak emerges as $h\to0$: The integrated conductivity is independent of $h$ (right inset) and equal to the Drude weight (left inset).}
\label{drude}
\end{figure}

Since for $h=0$ the heat current is conserved by the Hamiltonian, the AC conductivity $\sigma_h(\omega,h=0)=2\pi D\delta(\omega)$ features a Drude peak only. By generalizing Eq.~(\ref{Kubo}) to finite frequencies \cite{Sirkerprl,Sirker}, we can straightforwardly compute $\sigma_h(\omega,h)$ and demonstrate that it indeed becomes a $\delta$-function series for $h\to0$; the frequency-integrated heat conductivity just yields the Drude weight (`sum rule'). This is illustrated in Fig.~\ref{drude} and provides an independent test for the reliability of our extrapolation procedure.

{\it Bosonization}. We now present an analytic calculation of the low-temperature behavior of the charge conductivity using bosonization. We closely follow Refs.~\onlinecite{Sirkerprl,Sirker} which derive a parameter-free result for $\sigma_\textnormal{c}$ that is supposed to be correct if no conserved quantity protects the Drude weight. The current operator (\ref{current}) reads $J_c=-v\sqrt{K/\pi}\int dx\Pi$, and the Kubo formula for the AC conductivity reduces to $\sigma_c(\omega)=iK\omega\langle\phi\phi\rangle_r(\omega)/\pi$. The retarded correlation function $\langle\phi\phi\rangle_r$ is obtained by a perturbative field theory calculation to leading order in $h$, in $H_\textnormal{umklapp}$, and in $H_\textnormal{band curv.}$. The overall leading term governing the DC conductivity reads $\sigma_c=h^{-2}T^{3-2K}/C(K)$, which is consistent with Eq.~(\ref{T}) \footnote{The prefactor can also be obtained easily: $ C(K)=\pi^{K-3}(1-K^{-1}/2)^{2K-1}\cos^2(\pi K/2)\sin^{1-2K}(\pi K^{-1}/2) \Gamma^2(K/2) \Gamma^2(1/2-K/2)\Gamma^{2K}\left(\frac{1}{4K-2}\right)\Gamma^{-2K}\left(\frac{1}{2-K^{-1}}\right)\exp\Big(2\int_0^{+\infty}\frac{dx}{x}\Big(1-(K-1)e^{-2x}-\frac{\sinh x}{\sinh((K^{-1}-1)x)+\sinh x}\Big)\Big)$; bosonization predicts $\sigma_c\approx33$ at $\Delta=-0.85,h=0.2,T=0.05$, compared to the numerical result $\sigma_c\approx57$ (Fig. \ref{sigma}). Noting that $h=0.2$ is still not very small, the agreement is reasonable at all points in the regime $\Delta=-0.85,h=0.2,T\lesssim0.1$. and numerically we observe that $\partial\log\sigma_c/\partial\log h$ is slightly larger than $-2$ at $h=0.2$, implying that the agreement is better at smaller $h$.}. This term can be attributed to the staggered field, i.e.~the term which breaks integrability in the lattice model.

From the point of view of the field theory alone, the umklapp and staggered field terms have the same cosine structure; one of those terms is insufficient to break integrability of the \textit{field theory}. One might thus question the above reasoning and expect different temperature exponents in the conductivity depending on which of the two cosines was the `integrability-breaking small perturbation' on the integrable theory obtained by the other (i.e., different exponents for large and small $h$). As mentioned above, this picture is not supported by our DMRG data for the lattice model which agrees with the bosonization result over a broad range of values of $h$; however, we cannot rule out different behavior at even lower temperatures.

{\it Scaling analysis}. We finally carry out a simple scaling analysis. Combining several assumptions, which are likely to hold for other 1D models, we establish a scaling form for the conductivity in which all the exponents are determined up to one number, the scaling dimension of the integrability-breaking perturbation. For our model, this is consistent with the bosonization result (which also yields the scaling dimension).

It is reasonable to assume that Re$\langle J(t)J(0)\rangle/LT\approx A(\Delta,h,T)\exp(-\gamma(\Delta,h,T)t)$ at long time as correlations typically decay exponentially at finite $T$. The oscillation of this correlation function (see Fig. \ref{corr}) of Re$\langle J(t)J(0)\rangle/LT$ is not taken into account, as it cancels out in computing the integral (\ref{Kubo}). We also assume the amplitude $A(\Delta,h,T)$ does not vanish as $T\to0$ (note that the Drude weight $D(\Delta,h=0,T)$ is nonzero and continuous as $T\to0$). Then, (\ref{Kubo}) implies $\sigma_c\sim\gamma^{-1}$, and $\sigma_c$ takes the scaling form
\begin{equation}
\sigma_c(\Delta,h,T)=f(\Delta/T^{[\Delta]},h/T^{[h]})/T,
\label{scaling}
\end{equation}
where $[\Delta]$ and $[h]$ are the scaling dimensions of $\Delta$ and $h$, respectively. Note that $\sigma_c\sim T^{-1}$ for $[\Delta]=[h]=0$ or at the phase transition $\Delta=-\sqrt2/2,h=o(1)$.

In the perturbative regime (i.e., infinitesimal $h$), $[\Delta]=0$ as there is no renormalization of $\Delta$ (it is exactly marginal).  Then (\ref{scaling}) simplifies to $\sigma_c(\Delta,h,T)=f(\Delta,h/T^{[h]})/T$. As one expects $\sigma$ to diverge as $h^{-2}$ by a golden-rule argument \cite{Jungth} unless this perturbation is inefficient in inducing scattering, we take $f(x)\sim x^{-2}$.  Then
\begin{equation}
\sigma_c\sim h^{-2}T^{2[h]-1}=h^{-2}T^{3-2K}
\label{T}
\end{equation}
where in the second equality we have substituted the bosonization result $[h]=2-K$. Note that $\sigma_c\sim T^{-1}$ for $K=2$ or at the phase transition $\Delta=-\sqrt2/2,h=o(1)$.

The scaling analysis result is consistent with the bosonization calculation, which more convincingly justifies the assumptions in this section. However, it is worth pointing out that from the scaling analysis we still expect scaling of conductivity at low temperature in a gapless 1D system even when bosonization is inapplicable.  In general a gapless 1D system with a single velocity of low-energy excitations will be described by a conformal field theory (CFT) at long distances, and such theories are effectively ballistic as left-moving and right-moving excitations decouple.  We expect that the basic picture that conductivity is controlled by the leading irrelevant operator that induces scattering will still apply even when the CFT is more complicated than a free boson.

{\it Outlook}. Our work demonstrates an approach valid for many actual 1D materials, in which integrability-breaking terms are likely to be present but small. We studied one specific model but expect that our key result -- a power-law scaling of the conductivity $\sigma\sim T^{\alpha}$ with a universal exponent determined by the Luttinger liquid parameter -- should be a general qualitative feature of any gapless nonintegrable model in which no conserved operator protects the current. Quantitative results for other nonintegrable perturbations can be obtained by the numerical framework used in this paper. It should be possible to compute optical charge conductivities for comparison to experiments on conducting polymers and other systems.

On a more basic level, quantum critical transport in one dimension is controlled by the leading irrelevant operators if and only if those destroy integrability.  In higher dimensions, quantum critical transport is different because the critical theory is believed to be non-integrable, and transport properties are actively being studied by methods from high-energy physics.  Our results provide a constraint on these methods in a case where direct computation of transport coefficients is possible.

{\it Acknowledgements}: The authors thank F. Essler, F. Heidrich-Meisner, A. Rosch, J. Sirker, and M. Zaletel for useful comments and suggestions and acknowledge financial support from the Deutsche Forschungsgemeinschaft via KA3360-1/1 (C.K.), the Nanostructured Thermoelectrics program of LBNL (C.K.), the AFOSR MURI on ``Control of Thermal and Electrical Transport" (J.E.M.), and ARO via the DARPA OLE program (Y.H.).

\bibliography{scaling}

\end{document}